\documentclass[aip,jcp,amsmath,amssymb,reprint]{revtex4-1}
\usepackage{fancyhdr}
\fancypagestyle{legalnotice}{%
  \fancyhead[L]{}%
  \fancyhead[R]{}%
  \fancyfoot[C]{}%
  \fancyfoot[R]{\footnotesize Page \thepage\ of \pageref{LastPage}}%
  \fancyfoot[L]{\footnotesize The following article has been submitted to/accepted by the Journal of Chemical Physics.\\After it is published, it will be found at \url{https://aip.scitation.org/journal/jcp}.}%
}
\usepackage{graphicx}
\usepackage{dcolumn}
\usepackage{bm}

\usepackage[utf8]{inputenc}
\usepackage[T1]{fontenc}
\usepackage{mathptmx}
\usepackage{etoolbox}

\makeatletter
\def\@email#1#2{%
 \endgroup
 \patchcmd{\titleblock@produce}
  {\frontmatter@RRAPformat}
  {\frontmatter@RRAPformat{\produce@RRAP{*#1\href{mailto:#2}{#2}}}\frontmatter@RRAPformat}
  {}{}
}%
\makeatother

\usepackage{mathtools}
\usepackage{multirow}
\usepackage{siunitx}
\DeclareSIUnit{\basepair}{\text{bp}}
\usepackage{tikz}
\usepackage[textsize=small]{todonotes}

\newcommand{\kb}{k_\textnormal{B}}

\renewcommand{\vec}[1]{\mathbf{#1}}
\begin{document}
\title[A Numerical Investigation of Analyte Size Effects in Nanopore Sensing Systems]{
	A Numerical Investigation of Analyte Size Effects in Nanopore\\ Sensing Systems
}

\author{Kai Szuttor}
\email[]{kai@icp.uni-stuttgart.de}
\author{Patrick Kreissl}
\author{Christian Holm}
\email[]{holm@icp.uni-stuttgart.de}
\affiliation{Institute for
  Computational Physics, Universit\"at Stuttgart, Allmandring 3,
  D-70569 Stuttgart, Germany}

\date{\today}

\begin{abstract}
  We investigate the ionic current modulation in DNA nanopore
  translocation setups by numerically solving the electrokinetic
  mean-field equations for an idealized model.  Specifically, we study the dependence of
  the ionic current on the relative length of the translocating
  molecule.  Our simulations show a significantly smaller ionic current for
  DNA molecules that are shorter than the pore at low salt concentrations.
  These effects can be ascribed to the polarization of the ion cloud along the
  DNA that leads to an opposing electric dipole field.
  Our results for DNA shine light on the observed discrepancy between infinite pore
  models and experimental data on various sized DNA complexes.
\end{abstract}

\maketitle

\section{Introduction}
The basis of molecule sensing using nanopores is the modulation of the ionic
current in a setup where two electrolyte reservoirs are connected by a
nanopore.  An externally applied electric field causes an ionic current and
drives a charged particle or molecule through the nanopore. During the
translocation a salt-dependent ionic current modulation can be observed. Such
systems have been employed for a range of analytes from DNA and proteins to
viruses \cite{venkatesan11a, rotem12a, lewandowski19a}.  A recent study
\cite{taniguchi21a} even combined the nanopore sensing with machine learning to
differentiate different coronaviruses based on their footprint in the current
signal. Nanopore sensing has been thoroughly investigated to study the current
modulation caused by DNA molecules, both experimentally \cite{howorka01a,
fologea05a, storm05a, storm05b, smeets06a, keyser06a, keyser06b, clarke09a,
steinbock12a, keyser12a, bell16a, miles13a} and via numerical simulations
\cite{luan08a, keyser10a, getfert13a, lan11a, laohakunakorn13a, kesselheim14a, yeh12a, rempfer16a, rempfer16b, rempfer17a, weik16a,
weik19b, vandorp09a, chaudhry14a} using a multitude of simulation approaches\cite{aksimentiev10a}. A general overview of hydrodynamic
and transport phenomena in solid-state nanopores is given in several review articles
\cite{luo14a,fyta15a, ghosal19a}.  In theory such systems have often been investigated with
so-called infinite pore models \cite{kesselheim14a, weik16a, weik19b,
szuttor21a}. These models neglect any finite-length effects of the DNA molecule
or the pore by inherently assuming a translational symmetry.  For the
translocation of a single double-stranded DNA (dsDNA) molecule through a
cylindrical pore it has been shown \cite{weik19b} that infinite pore models
agree very well with the experimental data \cite{smeets06a,steinbock12a} for
the current modulation. While in these dsDNA nanopore systems the assumption of
an infinitely short pore holds, we want to explore the opposite limit in which
the analyte size approaches the length of the pore or is even shorter than the
pore. Therefore, we cover analyte-to-pore length ratios smaller than unity.
This range might, \textit{e.\,g.}, be relevant for DNA origami sensing systems
that incorporate glass nanocapillary pores with a sensing length of a few
hundred nanometers (\textit{cf.} Sec.~\ref{sec:pore_length} in the Appendix for
an estimating calculation) and an analyte with a similar or smaller length
\cite{wang19a}.

In this study, we present results for a finite pore model in terms of a
mean-field level description of the two-reservoir setup in which a DNA molecule is
fixed in the center of the pore. Our results show that the
ionic current through the pore significantly depends on the molecule length for
a range of electrolyte concentrations that typically contain the concentration
of zero modulation. Furthermore, we present an interesting effect of a large
current modulation when only one of the DNA ends is present in the pore. These
effects are expected to be much more noticeable for pores with a large sensing
length and cannot be resolved in typical DNA sensing setups with solid-state
nanopores due to the high demands on the ionic current sampling rate on the order of hundreds
of \si{\mega\hertz}.

Although we are not explicitly modeling a system of DNA origamis, the physical
mechanism, \textit{i.\,e.} the ion cloud polarization, leading to the observed
reduction in the ionic current is transferable to such systems. Future modeling
approaches in this direction based on the work presented here might also explain
the non-monotonic behavior of the cross-over salt concentration as a function of
the length of the origami molecule as observed by \citet{wang19a}.

\section{The DNA-Nanopore model}\label{sec:model}
\begin{figure}[ht]
    \centering
    \resizebox{\columnwidth}{!}{%
    \begin{tikzpicture}[every node/.style={scale=1.3, text=black}]
        \draw[thick, black] (-5, 0.5) -- (-0.5, 0.5) -- (-0.5, -0.5) -- (-5, -0.5); 
        \draw[thick, black] (5, 0.5) -- (0.5, 0.5) -- (0.5, -0.5) -- (5, -0.5);
        
        \begin{scope}
        \clip(-5.0, 0.5) rectangle (5,5.5);
        \draw[red, dashed, very thick] (0.0, 0.5) circle(5.0);
        \end{scope}
        \begin{scope}
        \clip(-5.0, -0.5) rectangle (5,-5.5);
        \draw[blue, dashed, very thick] (0.0, -0.5) circle(5.0);
        \end{scope}
        
        \draw[gray, -stealth, thick] (0.0, 0.0) -- (0.0, 3.0) node[above, xshift=-3.5]{$z$};
        \draw[gray, -stealth, thick] (0.0, 0.0) -- (3.0, 0.0) node[right]{$r$};
        
        \draw[gray, dashed] (0.0, -5.5) -- (0.0, 5.5);
        \draw[gray, -stealth, thick] (1.0, -3.0) arc(0.0:330:1.0 and 0.5);
        
        \draw[gray, stealth-stealth, thick](-3.0, -0.5) -- (-3.0, 0.5) node[right, pos=0.5]{$l_\mathrm{pore}$};
        
        \draw[red!50!white, fill=red!50!white] (-0.11, -2.0) rectangle (0.11, 2.0);
        \draw[red!50!white, fill=red!50!white] (0.0, 2.0) circle(0.11);
        \draw[red!50!white, fill=red!50!white] (0.0, -2.0) circle(0.11);
        \draw[gray, dashed] (0.0, 2.11) -- (-2.25, 2.25);
        \draw[gray, dashed] (0.0, -2.11) -- (-2.25, 2.15) node[text=black, above left, xshift=5.0]{$l_\mathrm{DNA}$};
        \draw[gray, very thick] (-0.11, 2.11) -- (0.11, 2.11);
        \draw[gray, very thick] (-0.11, -2.11) -- (0.11, -2.11);
        
        \node at (-3.0, -5.0) {$\Psi_-$};
        \node at (-3.0, 5.0) {$\Psi_+$};
        
        \node[anchor=west] at (1.0, -2.25) {cis reservoir};
        \node[anchor=west] at (1.0, 2.25) {trans reservoir};
        
        \draw[stealth-stealth, thick, gray] (-5.75, -5.5) -- (-5.75, 5.5) node[pos=0.5, anchor=center, right]{\rotatebox{90}{system length}};
        \end{tikzpicture}
        }
\caption{A sketch of the mean-field model of a nanopore with finite length.
The system is axisymmetric and can therefore be treated in two dimensions.
The hemispheric areas at the top and the bottom correspond to the cis and trans
electrolyte reservoirs.}
\label{fig:model}
\end{figure}
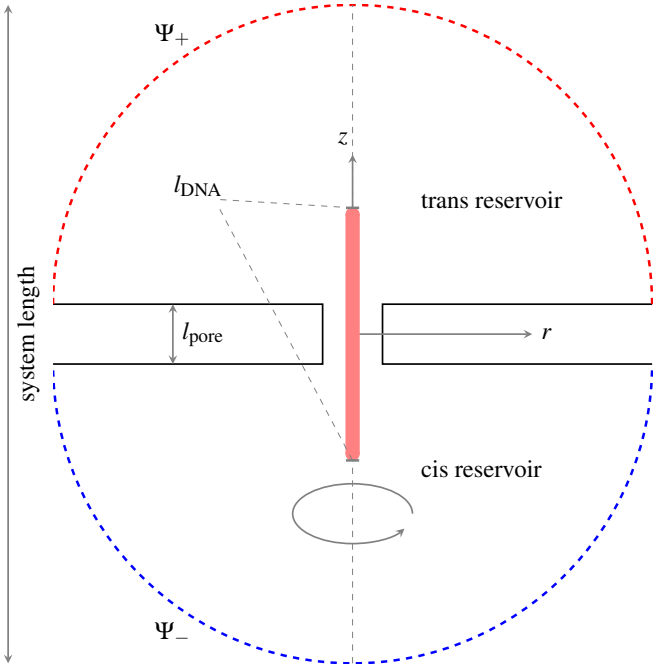

The model under study incorporates two electrolyte reservoirs with a monovalent
electrolyte solution of varying concentration that are connected by
a cylindrical nanopore. The DNA is modeled as a cylindrical object with rounded ends
fixed at the center of the pore. The DNA ends are modeled as hemispherical caps of the
same diameter as the DNA.
We mainly follow the modeling approach as described in Ref.~\onlinecite{weik19b}.
However, while Ref.~\onlinecite{weik19b} neglects any finite size effects of
the pore and the DNA molecule, the model in this work explicitly takes these
effects into account.
The sketch in Fig.~\ref{fig:model} shows the whole simulation
domain with the two electrolyte reservoirs, the connecting nanopore and the
DNA in the center of the pore. The diameter of the pore is \SI{10}{\nano\metre}
while the investigated pore length is \SI{40}{\nano\metre}. The length of the
DNA is varied in the range from 0.4 to 1.6 pore lengths (\textit{i.\,e.}
\SI{16}{\nano\metre} to \SI{64}{\nano\metre}) in order to also investigate the
conductivity for systems where the DNA is significantly shorter or larger than
the nanopore. The overall system length has been set to 10 times the pore length.

The mean-field model employed in this work is based on solving the
electrokinetic equations
(Poisson for electrostatics, Nernst-Planck for ion transport, and Stokes for hydrodynamics of the solvent)
for a charged cylinder representing the dsDNA in an
uncharged cylinder representing the nanopore following
the approach in Ref.~\onlinecite{weik19b}. The Nernst-Planck equation is
modified to incorporate the friction between ions and the dsDNA molecule that has
previously been found to be crucial for coarsened dsDNA models in order to
reproduce experimental and all-atom simulation data on current modulation in
nanopores~\cite{kesselheim14a, weik16a}.

The diffusion-advection equation for the fluxes of the ionic species $\vec{j}_\pm$ in the system reads:
\begin{align}
    \vec{j}_\pm = \left[-D\nabla+\vec{u}+\mu_\pm\vec{F}_\pm\right]c_\pm,
\end{align}
where $D_\pm$ is the diffusion constant, $\vec{u}$ is the fluid velocity,
$\mu_\pm=\mu=\frac{D}{\kb T} = \SI{4.8286e11}{\second\per\kilo\gram}$ is the ion mobility, $\vec{F}_\pm$ contains any external
forces, and $c_\pm$ the ion density. The external forces are comprised of the
electrostatic forces and the frictional forces:
\begin{align}\label{eq:visc_force}
    \vec{F}_\pm=ez_\pm \vec{E} -
\alpha \omega \left(0, 0, \frac{j^z_\pm}{c_\pm}\right)^\intercal.
\end{align}
Due to the velocity-dependent frictional force, this results in an algebraic
equation for the fluxes along the pore ($z$-component):
\begin{multline}
    j_{\pm}^z(r,z) = \left[-D\frac{\partial}{\partial_z} + u^z(r,z) \vphantom{\frac{j_{\pm}^z(r,z)}{c_{\pm}(r,z)}} +
                           \mu \left(\vphantom{\frac{j_{\pm}^z(r,z)}{c_{\pm}(r,z)}}e z_{\pm} E^z(r,z) \right. \right. -
                           \\\left. \left. \alpha \omega(r,z) \frac{j_{\pm}^z(r,z)}{c_{\pm}(r,z)} \right)\right]c_{\pm}(r,z),
    \label{eq:curr_dens}
\end{multline}
where $e$ is the
elementary charge, $z_\pm$ the valency of the ions, $E^z$ the electric field strength
along the symmetry axis, $\alpha=\SI{15e-12}{\kilo\gram\per\second}$ a numerical constant (controlling the
amount of frictional force) and $\omega$ is a position-dependent weight
function for the frictional force (adopted from Ref.~\onlinecite{weik19b}):
\begin{align}
\omega(r,z) =
\begin{cases}
	\left(1-\frac{r}{r_\mathrm{cut}}\right)^2 & \text{if } r<r_\mathrm{cut} \mathrm{~and~} |z|  < \frac{l_\mathrm{DNA}}{2}, \\
	0 & \mathrm{else,}
\end{cases}
\end{align}
where $r_\mathrm{cut}=\SI{1.4}{\nano\metre}$ is the cut-off distance for the
friction. We assume a temperature of \SI{300}{\kelvin} throughout all simulations.

The advective motion of ions is taken into account by means of Stokes' equations with incompressibility condition:
\begin{align}
	\eta \nabla^2 \vec{u} - \nabla p + \vec{f} &= 0,\\
	\nabla \cdot \vec{u} &= 0,
\end{align}
where $\eta$ is the dynamic viscosity, $\vec{u}$ the fluid velocity, $p$ the
pressure, and $\vec{f}=\sum_{i\in\{+,-\}}c_i \vec{F}_i$ (\textit{cf.}
Eq.~\eqref{eq:visc_force}) is the force density that the ions impose on the
fluid.
A no-slip boundary condition $\vec{u} = 0$ is applied on both the pore and DNA
surface. Hydrodynamic momentum exchange over the reservoir boundaries (dashed lines in
Fig.~\ref{fig:model}) is prevented by applying a vanishing normal stress condition for
the fluid flow.

Electrostatic interactions between ions are considered by solving the
Poisson equation 
\begin{align}
	\nabla^2 \Psi = - \frac{\sum_{i\in\{+,-\}} e z_i c_i}{\varepsilon_0 \varepsilon_\mathrm{r}},
\end{align}
where $\Psi$ is the electrostatic potential with vacuum permittivity
$\varepsilon_0$ and the relative permittivity of water
$\varepsilon_\mathrm{r}=78.54$\cite{wolf86a}. On the DNA boundary, we set the
surface charge density $\sigma = \SI{-0.136}{\coulomb/\meter\squared}$. For the
pore surface a zero-charge condition $\vec{n} \cdot \nabla \Psi$ = 0 is applied,
where $\vec{n}$ denotes the unit vector normal to the boundary. For the lower
and upper boundaries (drawn with blue and red dashed lines in Fig.~\ref{fig:model}, respectively) we set the
electrostatic potentials $\Psi_{-} = 0$ and $\Psi_{+} = l_\mathrm{pore}
E^z_\mathrm{ext}$ with $l_\mathrm{pore}$ being the length of the pore and $E^z_\mathrm{ext}=\SI{1e6}{\volt\per\meter}$ being the
approximate value for the electric field in comparable experimental setups
\cite{wang19a}. By choosing the electrostatic potential difference depending on
the pore length and thereby keeping the electric field in the pore approximately
constant we are able to directly compare results from different pores.
The friction between ions and dsDNA
results in a coupling between electrostatic and advective forces on the ions.
The model's inherent axisymmetry reduces the problem domain to two
dimensions. We solved the electrokinetic equations with the finite-element
method using the commercial software package COMSOL
Multiphysics\textsuperscript{\textregistered} version 5.6. (\textit{cf.}
Sec.~\ref{sec:fem_mesh} in the Appendix for information about the used FEM mesh).

\section{Results}
\begin{figure}[ht]
    \centering
    \includegraphics[width=\linewidth]{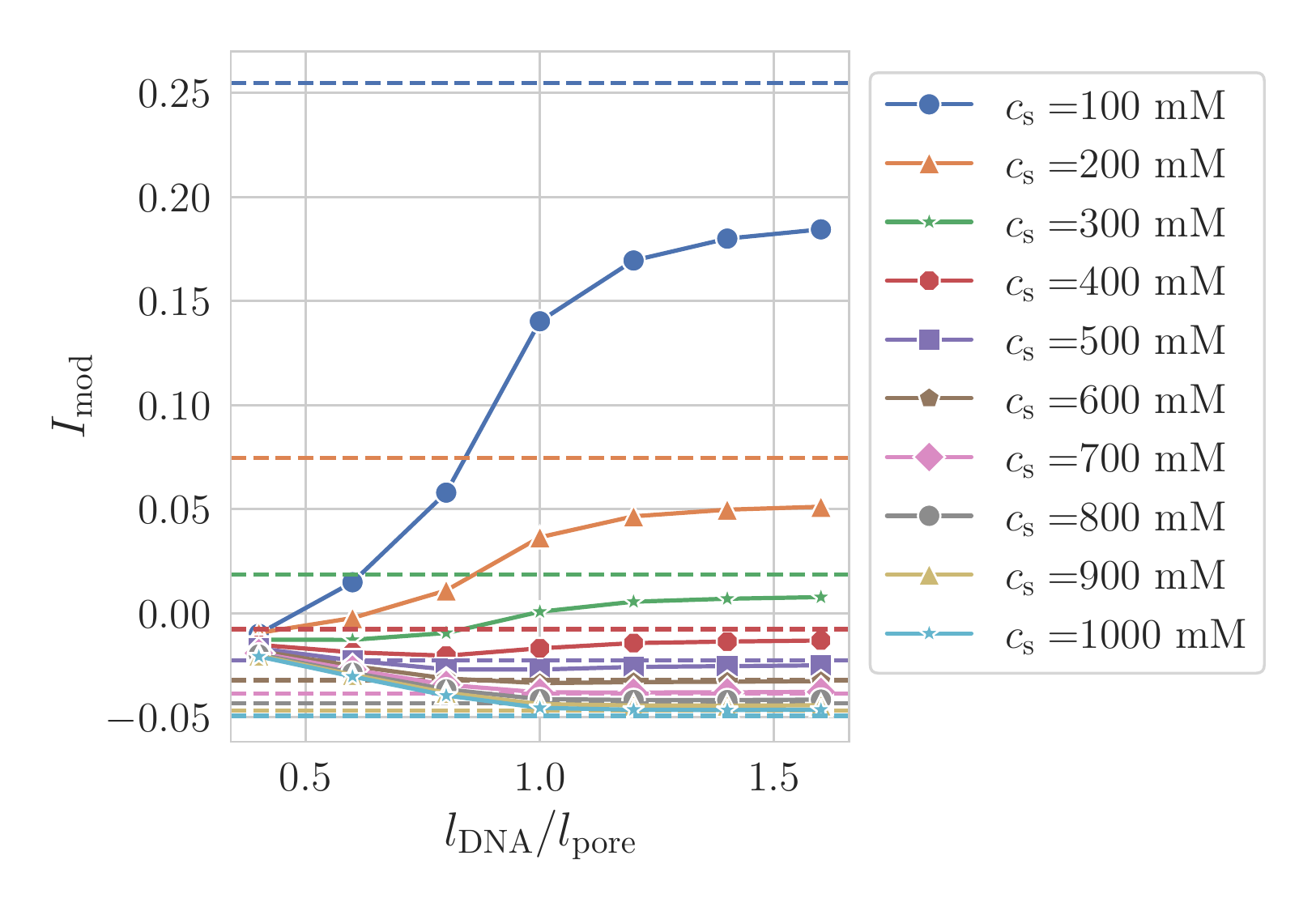}
    \caption{The ionic current modulation as a function of the DNA molecule's length
    for different salt concentrations. The length of the nanopore is
    \SI{40}{\nano\metre}. The dashed lines show the current modulation for an
    infinite pore (taken from results for the model as presented in Ref.~\onlinecite{weik19b}).}
    \label{fig:len_dep}
\end{figure}

The central physical quantity of interest is the ionic current through the DNA
nanopore system. In analogy to previous experimental and theoretical studies \cite{smeets06a,wang19a, weik19b,
szuttor21a}, we define the following pore length independent current modulation:
\begin{align}\label{eq:i_mod}
    I_\mathrm{mod}\coloneqq\frac{I_\mathrm{filled} - I_\mathrm{empty}}{I_\mathrm{empty}},
\end{align}
where $I_\mathrm{filled}$ is the ionic current through the pore if a test
molecule is present in the pore and $I_\mathrm{empty}$ is the empty pore
current.
In Fig.~\ref{fig:len_dep}, the ionic current modulation is shown  as a function of the
DNA length for different electrolyte bulk concentrations. 
There is a significant drop in the modulation for shorter DNA molecules if the
molecule length falls below the pore length.

\begin{figure}[ht]
    \centering
    \includegraphics[width=\linewidth]{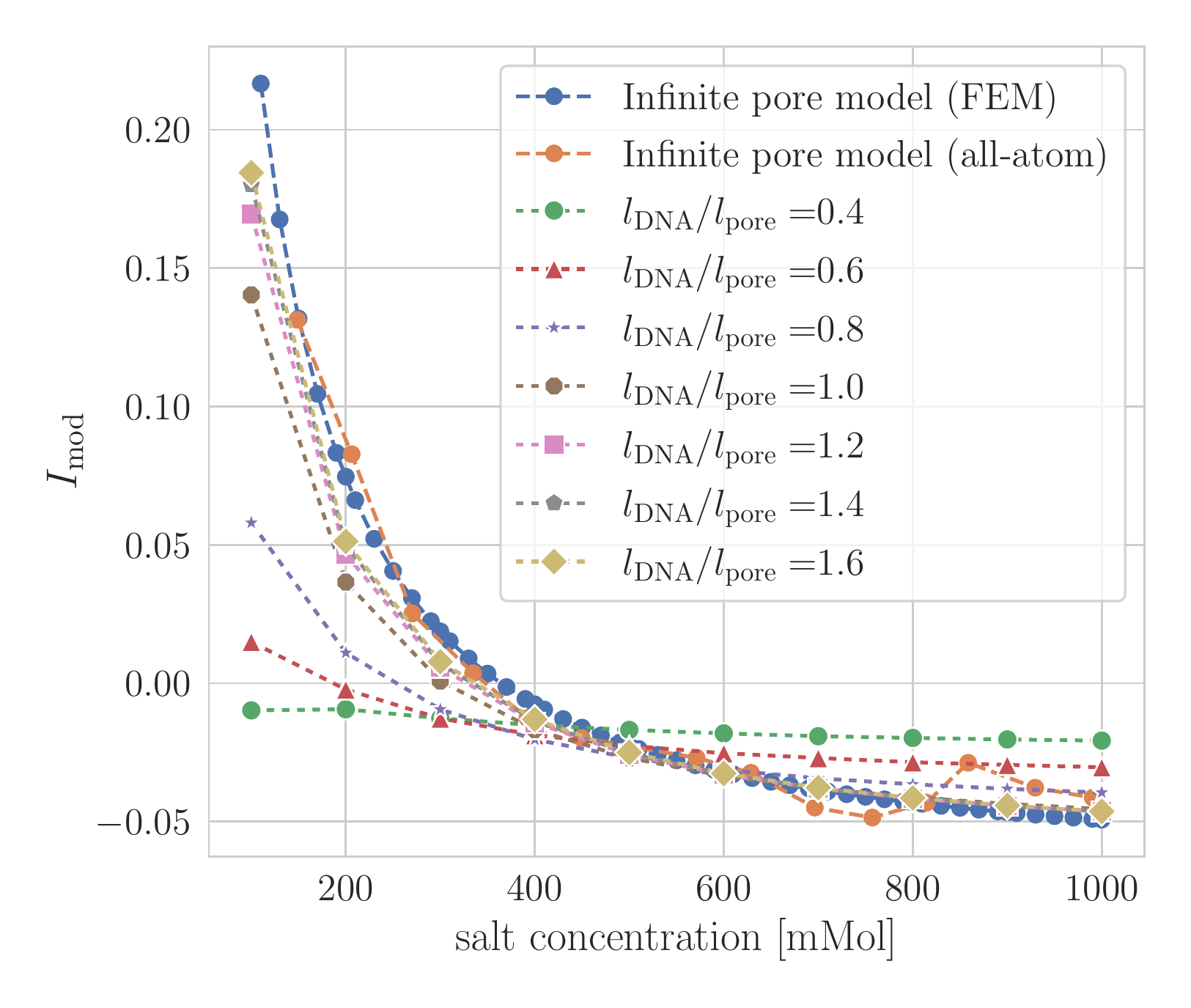}
    \caption{The ionic current modulation as a function of salt concentration
    different DNA molecule lengths. The data is also compared to results from two
    infinite pore models on the mean-field level~\cite{weik19b} and the
    atomistic level of detail~\cite{kesselheim14a}.}
    \label{fig:compare_to_inf}
\end{figure}

\begin{figure}[ht]
    \centering
    \includegraphics[width=\linewidth]{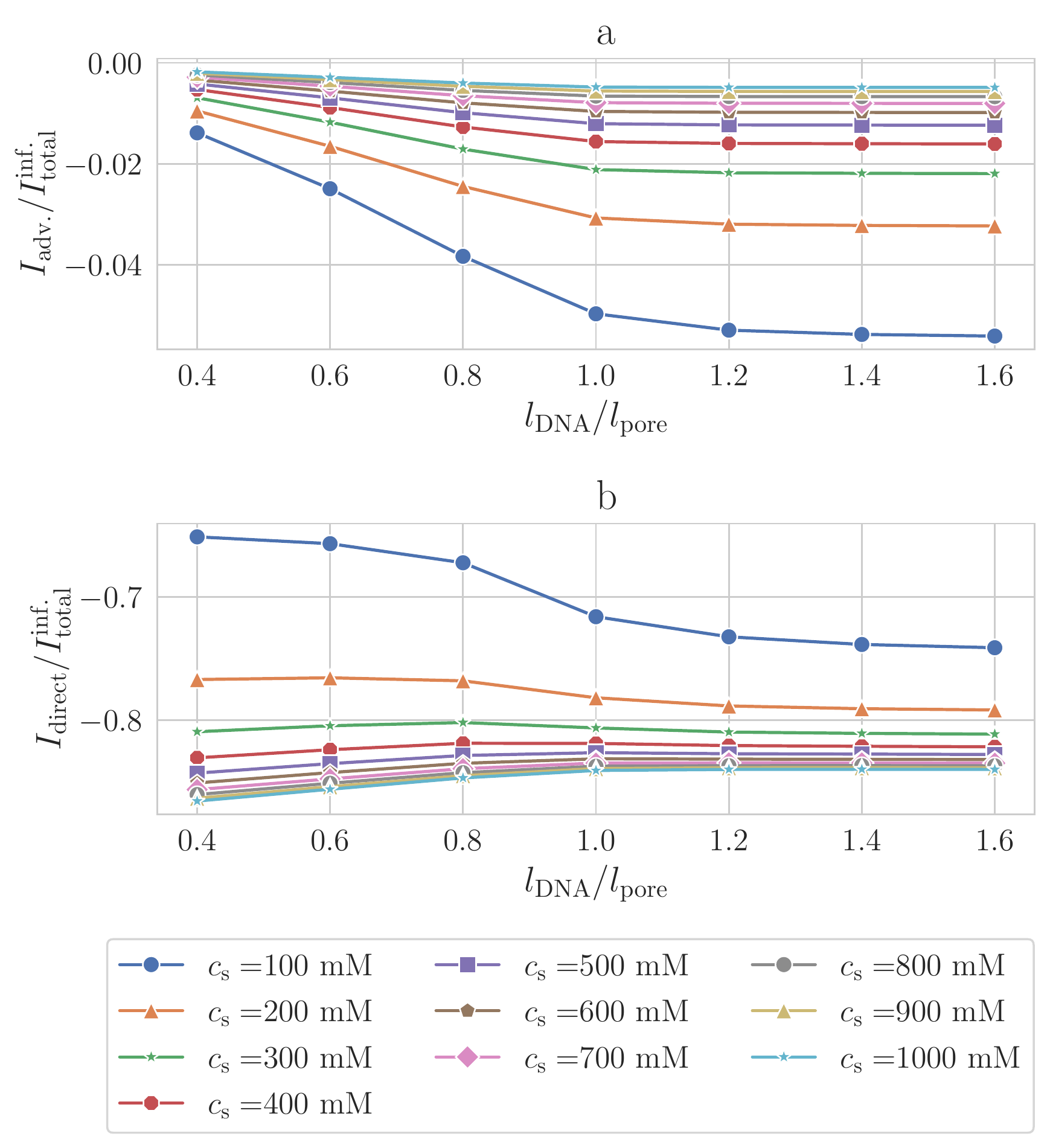}
    \caption{The advective (a) and direct (b) contributions to the ionic current
    as a function of the DNA length.}
    \label{fig:current_components}
\end{figure}

Comparing the current modulation to data from the infinite pore model
of Ref.~\onlinecite{weik19b} (\textit{cf.} Fig.~\ref{fig:compare_to_inf}), the
curves for the finite pore seem to converge towards the infinite pore reference
for increasing ratios of $l_\mathrm{DNA}/l_\mathrm{pore}$.
The differences in the modulation are larger for shorter molecules and smaller
salt concentrations.
Consistent with results from all-atom simulations of an infinite pore system as
presented in Ref.~\onlinecite{kesselheim14a}, the ionic current through the pore
is dominated by the direct contribution caused by the applied electric field
(\textit{cf.} Eq.~\eqref{eq:direct_current_density}). As can be seen in
Fig.~\ref{fig:current_components}a, the advective current fraction due to
electroosmotic flow along the DNA is at most in the
order of 5--6 percent. Notably, the ionic current's dependency on the DNA length
has its maximum for the lowest salt concentration, \textit{i.\,e.} where the negatively-charged
surface is screened least. This is in line with the salt-dependent ionic current
modulation in experimental systems \cite{smeets06a,wang19a} and simulation studies
\cite{kesselheim14a, weik19b, szuttor21a} of similar systems in which the current modulation diverges in the limit of zero bulk electrolyte concentrations.

Because of the much larger influence of the direct current on the total current modulation, we focus further analysis on the electric field and the amount of charge in the pore since these
quantities actually define the direct current density (neglecting constants):
\begin{align}\label{eq:direct_current_density}
    j^z_\mathrm{direct} (r,z)= \frac{\mu e^2  E_z(r,z)  \left[c_+(r,z) + c_-(r,z)\right]}{\alpha  \mu  \omega(r,z) + 1},
\end{align}
where we reused the notation of Eq.~\ref{eq:curr_dens}.

\begin{figure}[ht]
    \centering
    \includegraphics[width=\linewidth]{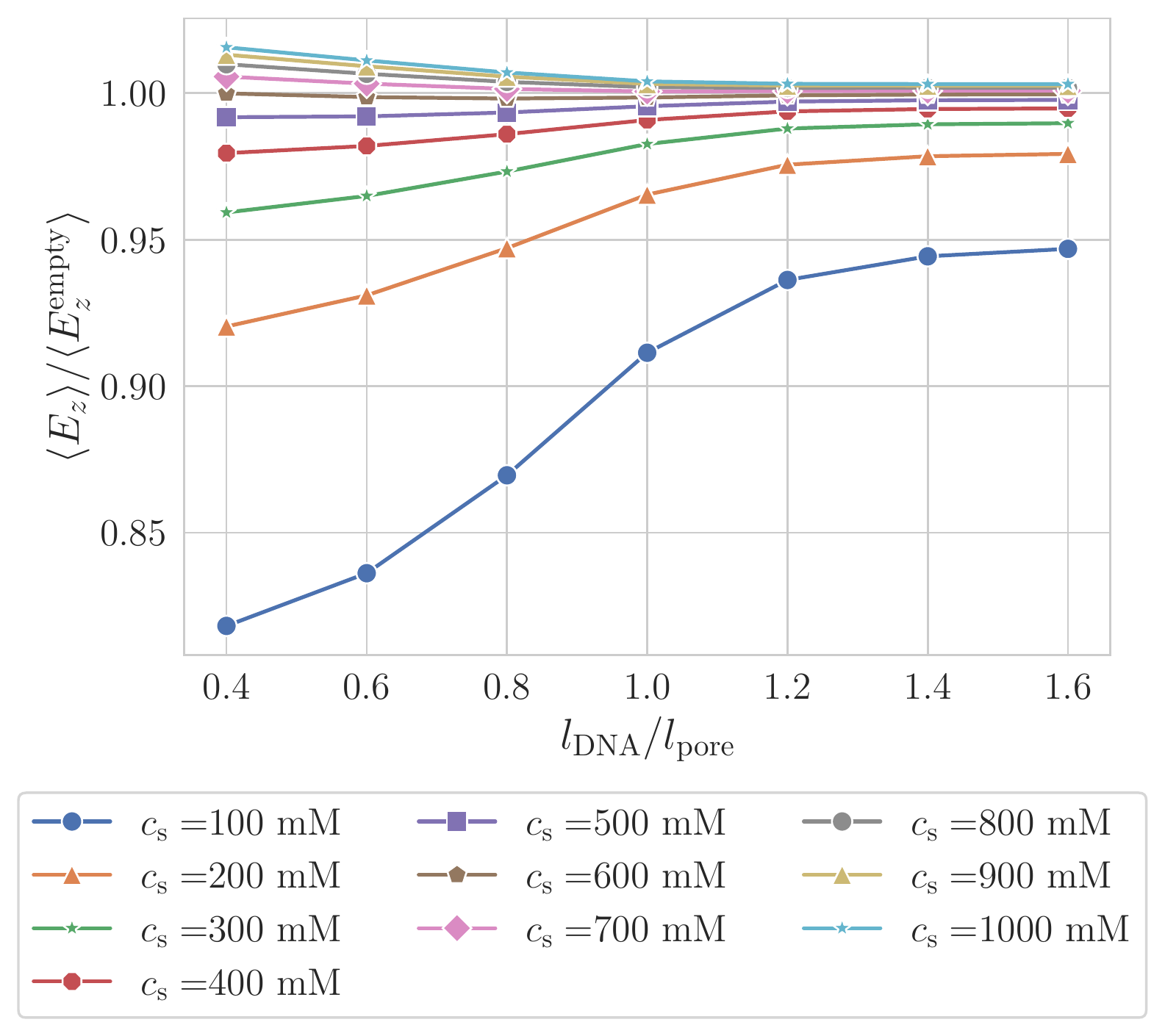}
    \caption{The average $z$-component of the electric field in the center of the pore as a function of the DNA
    length normalized by the average $z$-component of the electric field in the center of an empty pore.}
    \label{fig:electric_field}
\end{figure}

In order to investigate the electric field's dependency on the DNA length we
radially averaged the $z$-component of the field at the center of the pore
($z=0$ and $r\in \left[ r_\mathrm{DNA}, r_\mathrm{pore}\right]$).
Along this horizontal line, the electric field vanishes for symmetry reasons if
no external field acts on the mobile charges of the system. However, as can be
seen in Fig.~\ref{fig:electric_field} if an
external field is applied we observe a significant modulation of the
electric field along the pore as a function of the DNA length for small
electrolyte concentrations. In the most intuitive picture on the level of an
equivalent circuit model for the DNA nanopore system the DNA acts as an increased resistance in the pore
and the total resistance of the DNA increases with its length. Thus, a larger portion
of the potential drops along the DNA which increases the electric field in the
pore. However, as Fig.~\ref{fig:len_dep} shows, the DNA actually enhances the
current in the pore (for bulk electrolyte concentrations up to around \SI{400}{\milli\mole\per\litre}). Therefore,
this simplified picture can not explain our observation of an increased electric
field.

\begin{figure}[ht]
    \centering
    \includegraphics[width=\linewidth]{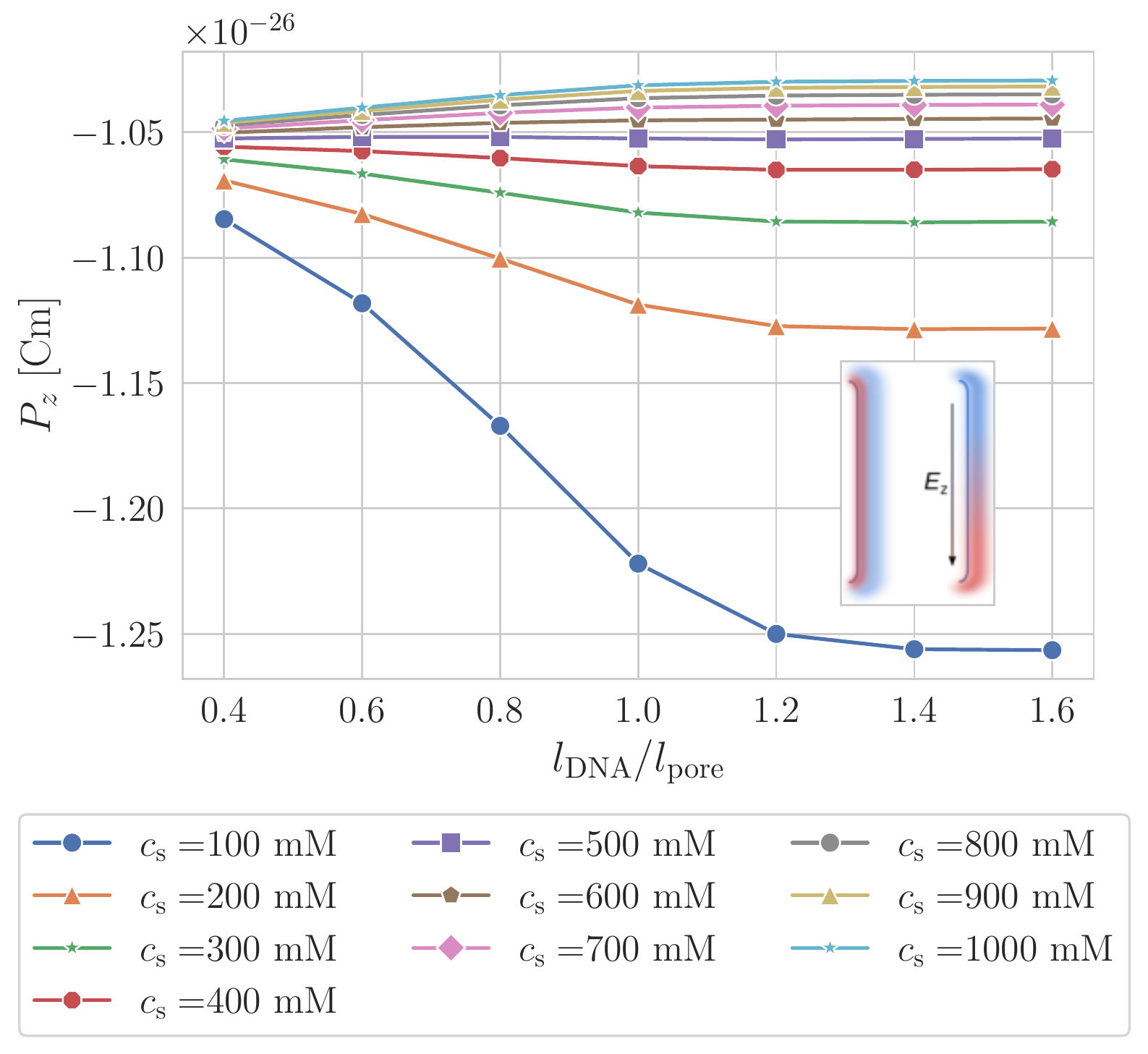}
    \caption{The $z$-component of the total electric dipole of the whole
    simulation domain. The inset image shows a sketch of the ion cloud around
    the charged surface of the DNA: the external field induces an asymmetry in
    the mobile ion distribution.
    }
    \label{fig:dipole_moment}
\end{figure}

The external electric field induces an electric dipole field caused by the polarized counter-ion cloud
around the negatively charged DNA that weakens the externally
applied field in the pore. Fig.~\ref{fig:dipole_moment} shows the dependency of this dipole
moment on the DNA length. Here, again we see a significant dependency on the DNA
length for low salt concentrations. In addition, the induced dipole moment
decreases with the salt concentration. Intuitively one might think that a larger
dipole moment would lead to a larger opposing field and therefore a weaker total
field in the pore center. The dipole electric field in the pore center, however,
also depends on the locations of the charges.

In the following, we want to develop a simple model to get a better
understanding of what the induced dipole moment's data actually means for the
electric field in the pore. In this model we assume that all electrostatic
effects of the mobile ions and the negatively charged DNA are included in the
resulting electric dipole field. We therefore calculate the electric field of
two charges that are located at the two ends of the DNA on the symmetry axis of
the system at ($\pm 0.5 l_\mathrm{DNA}$). This results in an electric field in
the center of the pore (\textit{cf.} Sec.~\ref{sec:dipole_model} in the Appendix):
\begin{align}
    E_z^\mathrm{dipole}(r, z=0) = \frac{2\left|P_z\right|}{\pi\varepsilon_0\varepsilon_\mathrm{r}\left(l_\mathrm{DNA}^2+4r^2\right)^{\frac{3}{2}}},
\end{align}
where $P_z$ is the induced dipole moment, $\varepsilon_0$ is the vacuum
permittivity and $\varepsilon_\mathrm{r}$ is the permittivity of water.
The resulting electric field radially averaged over the pore at $z=0$ decreases with increasing DNA
length (\textit{cf.} Fig.~\ref{fig:dipole_electric_field} in the Appendix) which
explains the observation of an increasing total electric
field in the pore center: the decreasing electric field strength of the induced dipole
has less influence on the total electric field for longer DNA molecules. In
addition, there is a reduced influence of the electric field on the DNA length
for increasing electrolyte concentrations. From $c_\mathrm{s}\approx
\SI{700}{\milli\mole\per\litre}$ on the DNA charge seems to be completely
screened. For these high salt concentrations the DNA can be seen as an
additional resistor, therefore causing a higher portion of the electrostatic
potential to drop along the DNA which is equivalent to a larger electric field strength.

\begin{figure}[ht]
    \centering
    \includegraphics[width=\linewidth]{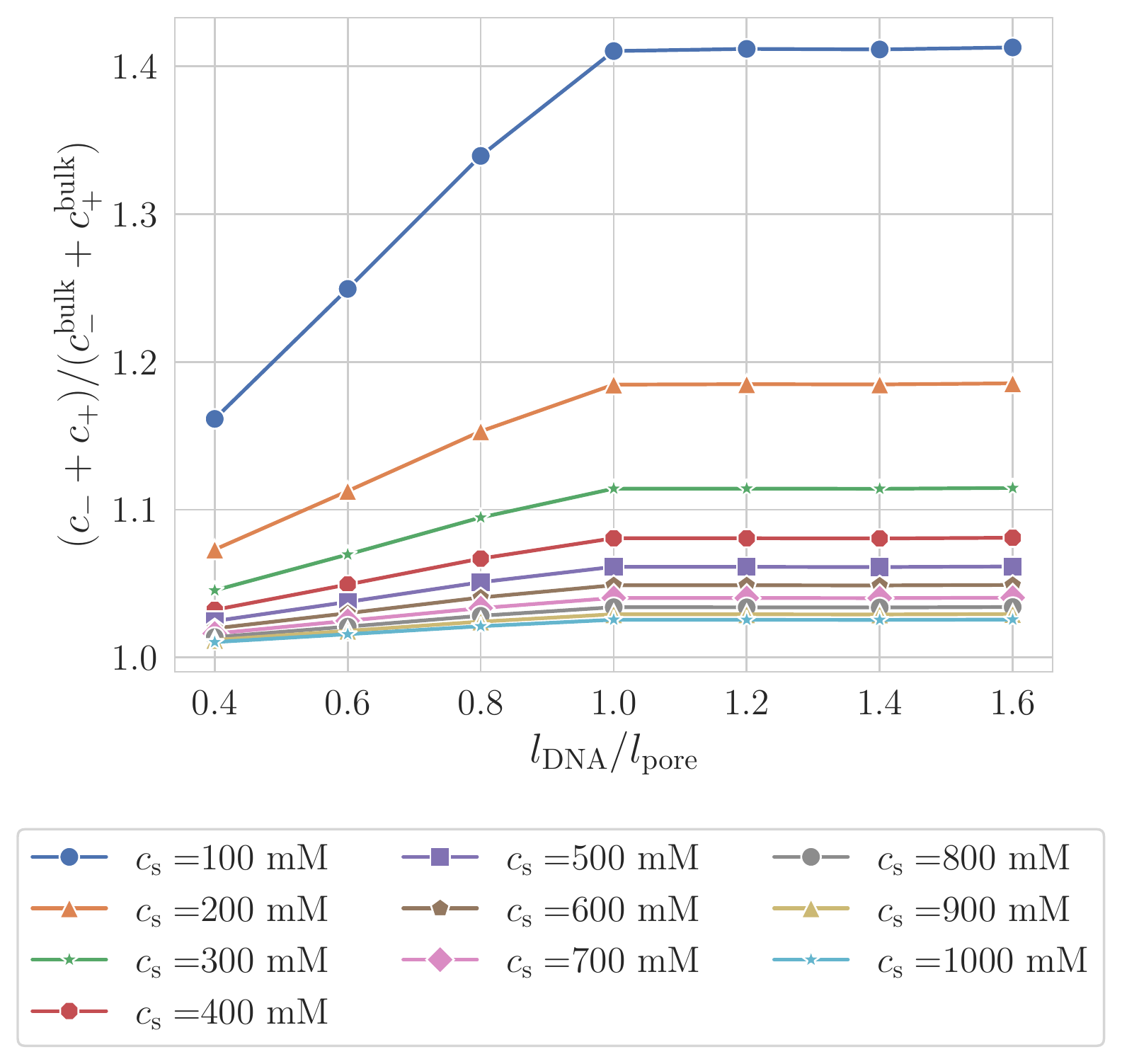}
    \caption{The averaged total ion density in the pore normalized with the respective bulk values as a function of the DNA length.}
    \label{fig:total_ion_density}
\end{figure}

The second contributing factor to the direct current is the total
charge in the pore. Therefore, we analyze the DNA length's influence on the
ion density in the pore. In the pore, the presence of the
negatively charged DNA repels the co-ions and attracts the
counter-ions. In addition, this effect is enhanced for longer DNA molecules
(\textit{cf.} Sec.~\ref{sec:ion_density} in the Appendix).
The total ion density in the pore, however, is enhanced (compared to the reservoir) and increases
with the length of the DNA. Therefore, the repulsion of co-ions is
overcompensated by the attraction of the counter-ions (\textit{cf.} Fig.~\ref{fig:total_ion_density}). This observation is in
line with the overall observation of a larger ionic current through the pore for
electrolyte concentrations of up to about \SI{400}{\mole\per\litre}.

\begin{figure}[ht]
    \centering
    \includegraphics[width=\linewidth]{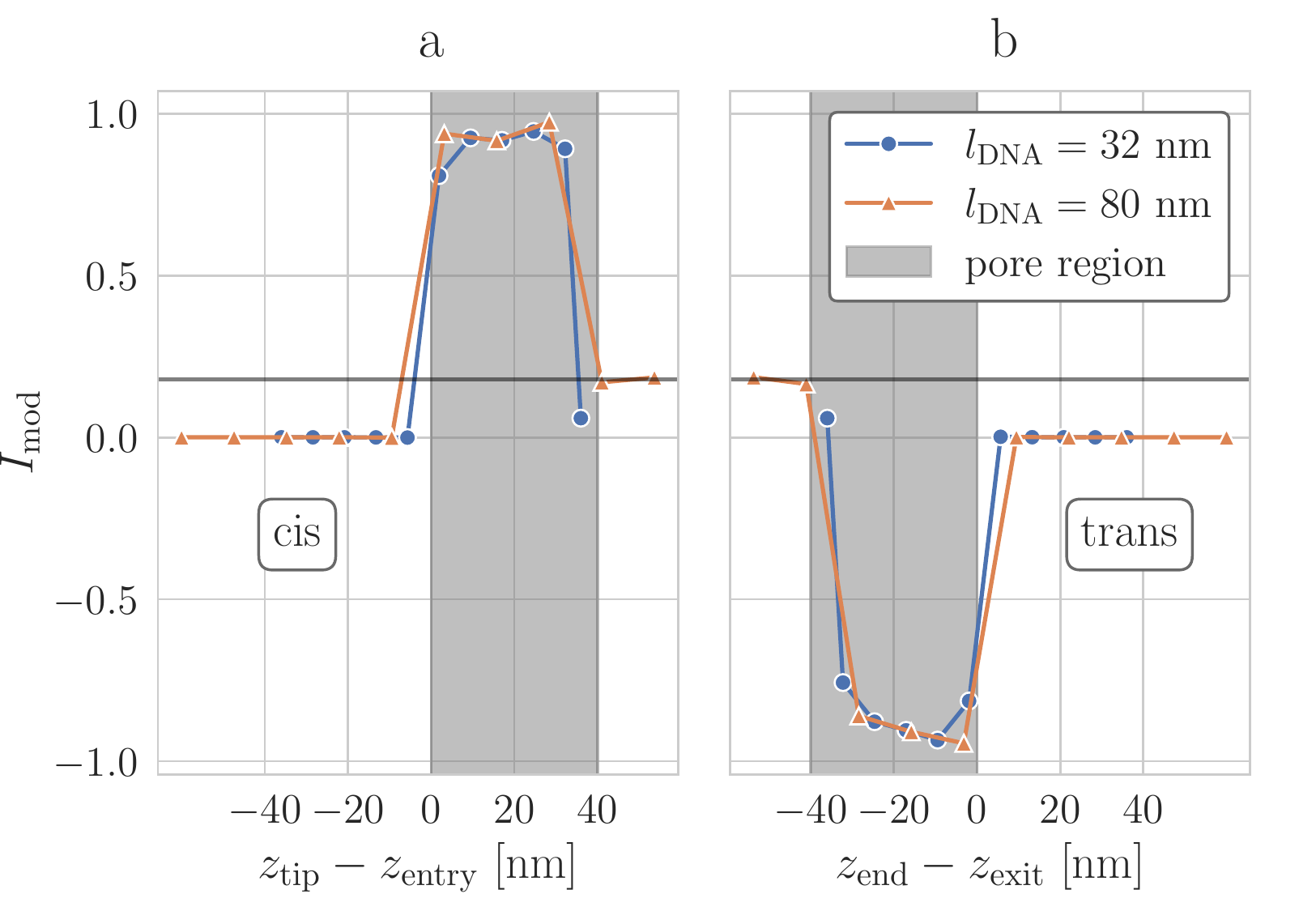}
    \caption{The ionic current modulation (\textit{cf.} Eq.~\eqref{eq:i_mod})
    as a function of the distance between the DNA tip and the pore entry (a) and
    as a function of the distance between the DNA end and the pore exit
    (b) for two different DNA lengths. The horizontal line shows the approximate
    modulation when both DNA ends are outside of the pore region.}
    \label{fig:varying_dna_pos}
\end{figure}

Fig.~\ref{fig:varying_dna_pos} shows an interesting effect of an enhanced ionic
current modulation that we observe in the two phases of the translocation
process in which only one end of the DNA is located in the pore.
Notably, this modulation is about a factor of four larger compared to the
situation where neither of the DNA ends is in the pore. This high
modulation is only observed for a relatively short part of the translocation of
about \SI{20}{\nano\metre} in our pore geometry and therefore the sampling
frequency of experimental setups would need to be sufficiently large to
capture such effects. The translocation speed of a DNA through a solid state
nanopore is about $\SIrange{1e6}{1e8}{\basepair\per\second}$, \textit{i.\,e.}
$\SIrange{0.34}{34.0}{\milli\metre\per\second}$ ~\cite{plesa15a, heerema18a}. That means that
the time interval that has to be resolved in the experiment is about $\Delta
t\approx\SIrange{0.588}{5.88}{\micro\second}$ which translates to a sampling rate of about 
$\SIrange{1.7}{17}{\mega\hertz}$.
However, since the high modulation is present as long as only one of the analyte
ends is in the sensing region, it might be possible to resolve the modulation
for pores with a longer sensing length, \textit{e.\,g.} in glass nanocapillaries.
Still, we are not aware of experimental observations of similar effects.

\section{Conclusion}
We have shown that the current modulation of analytes in nanopore sensing setups
significantly depends on the length of the translocating charged object if the
length of the object is shorter than the sensing length of the pore for
electrolyte concentrations up to about \SI{400}{\milli\mole\per\litre}.
Our model in this work extends a previously studied mean-field model
\cite{weik19b} for an infinite dsDNA molecule in an infinite pore to a model
that includes a finite pore and the cis- and trans-reservoirs of the
Coulter-type sensing setup. Our data shows that the ionic current increases with
the length of the DNA and converges to values of infinite pore models as
presented in Refs.~\onlinecite{kesselheim14a, rau17a, weik19b, szuttor21a}.
Furthermore, we investigated the underlying electrostatic effects,
namely the length-dependent opposing electric field that results from
the polarized ion cloud around the DNA. This dipole field's component
parallel to the pore gets smaller for an increasing DNA length which leads to an
overall increasing electric field in the pore. Such effects are not present
our infinite pore models for DNA origamis as presented in
Ref.~\onlinecite{szuttor21a} which might explain the deviation between the
simulation data of these models and experimental results of \textcite{wang19a}.

In addition, our simulations revealed an interesting effect, namely
the appearance of an enhanced ionic current modulation upon just
entering or just leaving the nanopore interior that can be ascribed to
the local electric field of the charged molecule's tip either
amplifying or weakening the externally applied electric field in the
pore.

\begin{acknowledgments}
We thank Florian Weik, Alexander Schlaich, Jonas Landsgesell and Georg Rempfer
for helpful and constructive discussions.
The work is funded by the Deutsche Forschungsgemeinschaft (DFG, German Science
Foundation) — Project Number 390740016 — EXC 2075 ``SimTech''.
\end{acknowledgments}

\section*{Data Availability Statement}
The data that support the findings of this
study are available from the corresponding author upon reasonable request.

\appendix

\section{Effective Pore Length} \label{sec:pore_length}
\begin{figure}[ht]
	\centering
	\includegraphics[width=\linewidth]{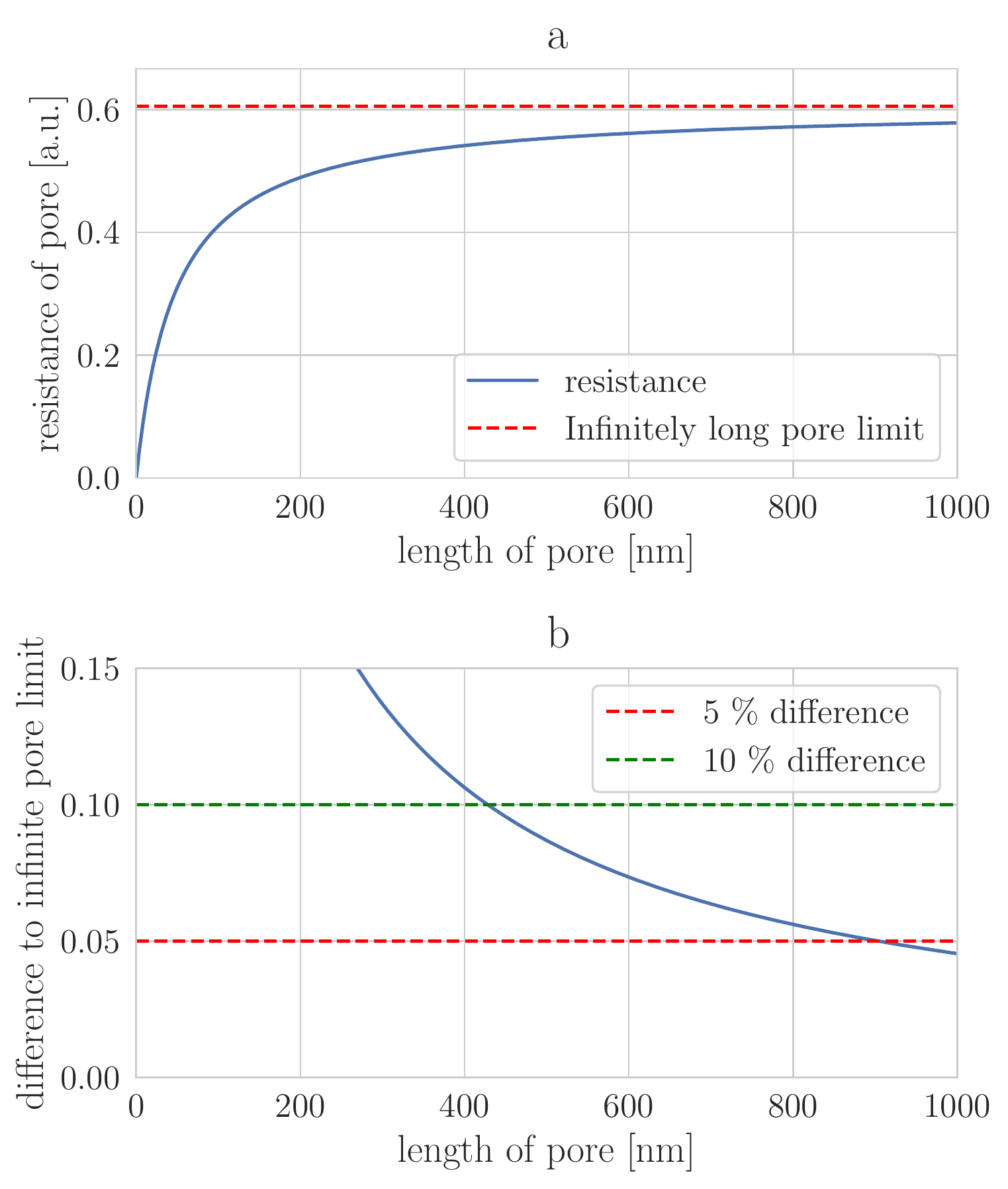}
	\caption{Length-dependent pore resistance for a conical pore and the
		respective resistance in the infinite pore limit (a). The difference between the
		resistance of the conical pore with a given length and the infinite pore
		limit (b). The crossing of the \SI{5}{\percent} line and the resistance curve is
		at about \SI{900}{\nano\metre}.}
	\label{fig:conical_res}
\end{figure}

In experimental studies of \citet{wang19a} a similar setup to the model presented in this manuscript has been utilized
to investigate the current modulation of complex DNA structures. One
of the more significant differences, however, is the pore geometry. The glass
capillaries used in the experimental study have a diameter of about
\SI{10}{\nano\metre} at the tip but have a conical geometry with an opening
angle of about \SI{6}{\degree}. It is known from experimental~\citep{vandorp09a} as well as
simulation studies~\citep{weik19b} that a wider nanopore causes a larger
magnitude of electroosmotic flow which leads to a larger contribution of
the advective current to the overall ionic current. However, all-atom simulations in an infinite
cylindrical pore setup~\citep{kesselheim14a} revealed that this contribution is
negligible for infinite DNA nanopore systems. Nevertheless, even if this
geometric effect might not be significant for the conductivity of the system,
the absolute length of the nanopore might influence the importance of
the finite size effects. To get an estimate for up to which length the conicity of the
experimental pore affects the conductivity, we calculated the length-dependent
pore resistance. For an infinitesimal pore segment of length $\mathrm{d}z$ the resistance of a
conical pore with the cross-section area $A(z)$ can be described as
\begin{align}
	\mathrm{d}R = \rho \frac{\mathrm{d}z}{A(z)},
	\label{eq:diff_res}
\end{align}
where $\rho$ is the bulk electrolyte resistance and $A(z)= \pi r(z)^2 = \pi (r_1
+ z\tan \alpha)^2$ is the cross-section area of the pore at the position
$z$ measured from the tip of the pore. Here, $\alpha$ denotes the opening angle
of the conical pore geometry. Integrating Eq.~\ref{eq:diff_res} yields:
\begin{align}
	R(z) = \rho \int_0^z \mathrm{d}\tilde{z} \frac{1}{A(\tilde{z})} = \frac{\rho}{\pi} \frac{z}{(r_1 (z+\frac{r_1}{\tan \alpha}))\tan \alpha }
	\label{eq:conical_res}
\end{align}
In Fig.~\ref{fig:conical_res}, Eq.~\ref{eq:conical_res} is shown for a pore
length up to \SI{1}{\micro\metre}. The value of the bulk resistance
$\rho$ is arbitrarily set to unity.
In the limit of an infinitely long conical pore Eq.~\eqref{eq:conical_res}
converges to a resistance of $R_\infty = \frac{\rho}{\pi\tan(\alpha)r_1}$.
However, as
shown in Fig.~\ref{fig:conical_res}, the largest contribution to the resistance
(\SI{95}{\percent} at about \SI{900}{\nano\metre}) suggests an effective
length of the
pore on the order of several hundred nanometers. Although this model only takes
into account the resistance based on the geometry of the pore, the result is in
line with simulation studies in Ref.~\onlinecite{bell16a} in which the decay of the
electric field in the pore was used to estimate the sensing length.

The DNA structures studied by Wang \textit{et\,al.} in Ref.~\onlinecite{wang19a} are
between approximately \SI{150}{\nano\metre} and \SI{600}{\nano\metre} and
are therefore shorter than the estimated effective pore length. This clearly
sets those DNA nanopore systems apart from many other experimentally
investigated setups in which the DNA is orders of magnitude longer than
the pore.

\section{Ion Density in the Pore}\label{sec:ion_density}
\begin{figure}[h!]
	\centering
	\includegraphics[width=\linewidth]{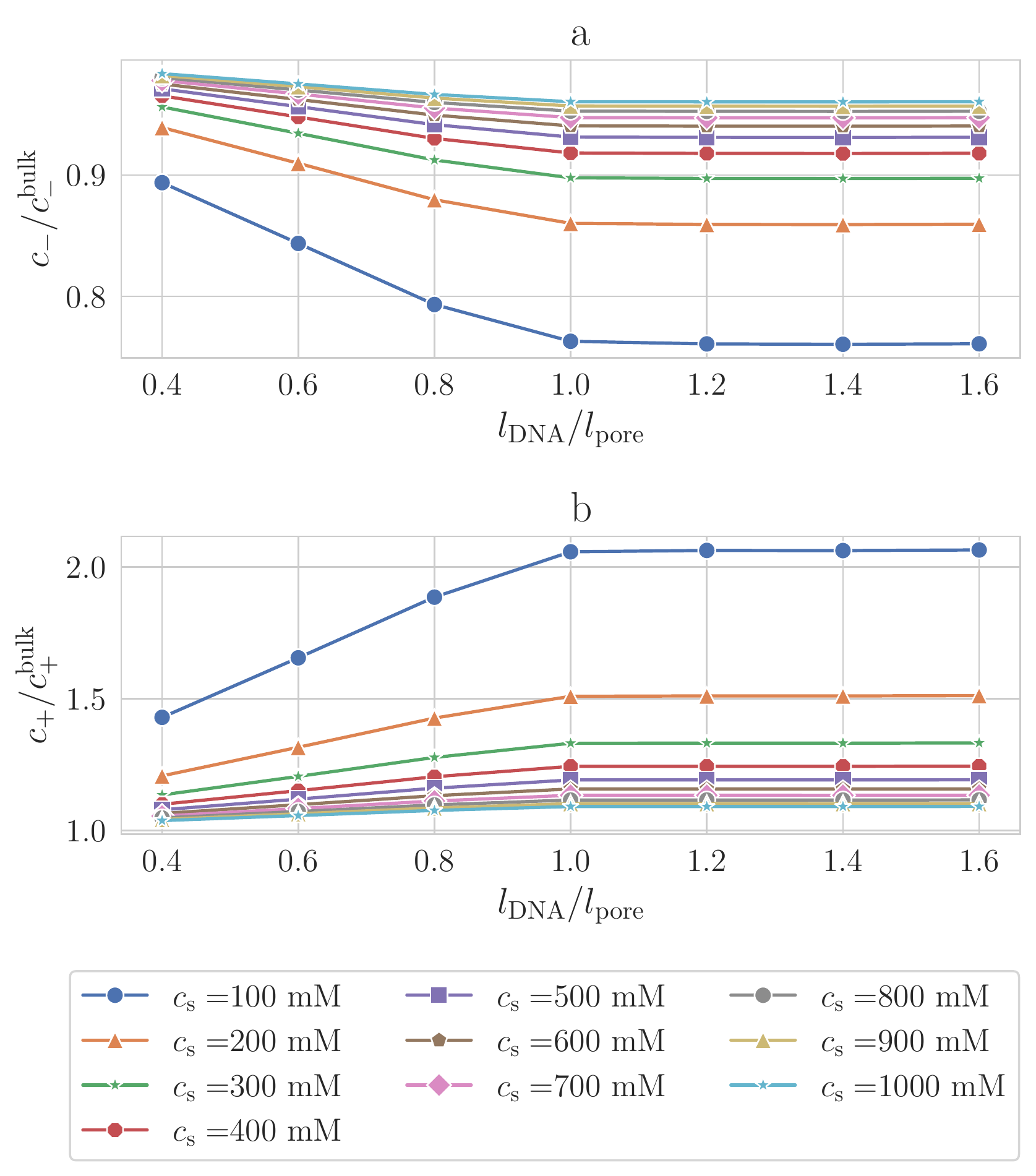}
	\caption{The average anion (a) and cation density (b) in the pore normalized with the respective bulk values as a function of the DNA length.}
	\label{fig:ion_density_in_pore}
\end{figure}

We analyzed the density of the two ion species in the pore.
The co-ions of the dsDNA are repelled from the pore and the density is well
below the bulk value for all investigated DNA lengths and salt concentrations.
The cation density, however, is enhanced in the pore and increases with the
length of the DNA (\textit{cf.} Fig.~\ref{fig:ion_density_in_pore}).

\section{Electric Dipole Field Model}\label{sec:dipole_model}
\begin{figure}[h!]
		\centering
		\includegraphics[width=\linewidth]{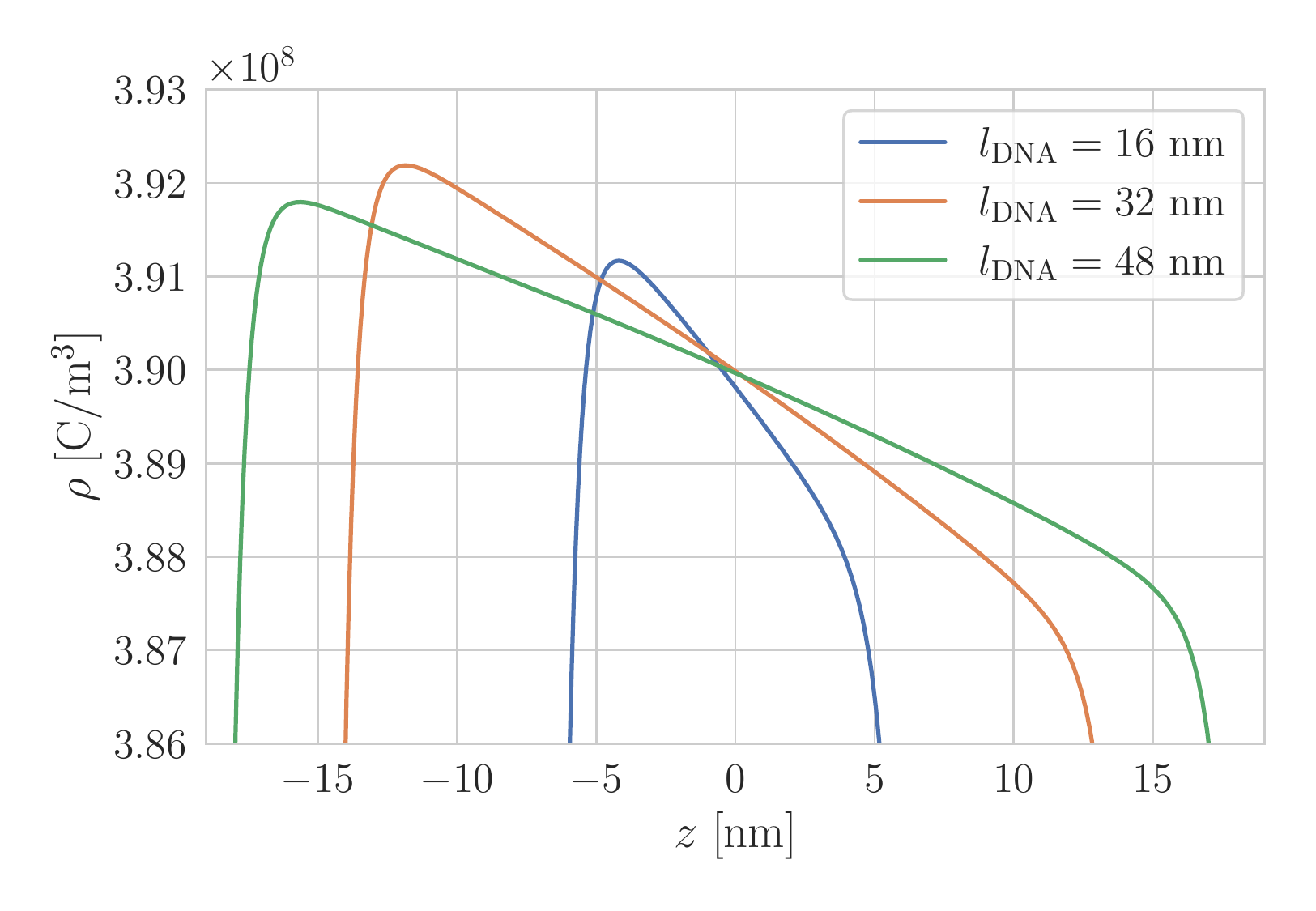}
		\caption{The charge density along the $z$-axis at $r=r_\mathrm{DNA}$ for
		three different DNA lengths. The ion cloud at the vicinity of the DNA is
		asymmetric due to the externally applied electric field.}
		\label{fig:charge_density_along_DNA}
\end{figure}

This model assumes that the $z$-component of the total dipole moment DNA nanopore
system is created by a single pair of charges at the two ends of the DNA on the
symmetry axis. This assumption is based on the observation that the
charge density is asymmetric along the DNA as shown in
Fig.~\ref{fig:charge_density_along_DNA}.
\begin{figure}[h!]		
		\centering
		\includegraphics[width=\linewidth]{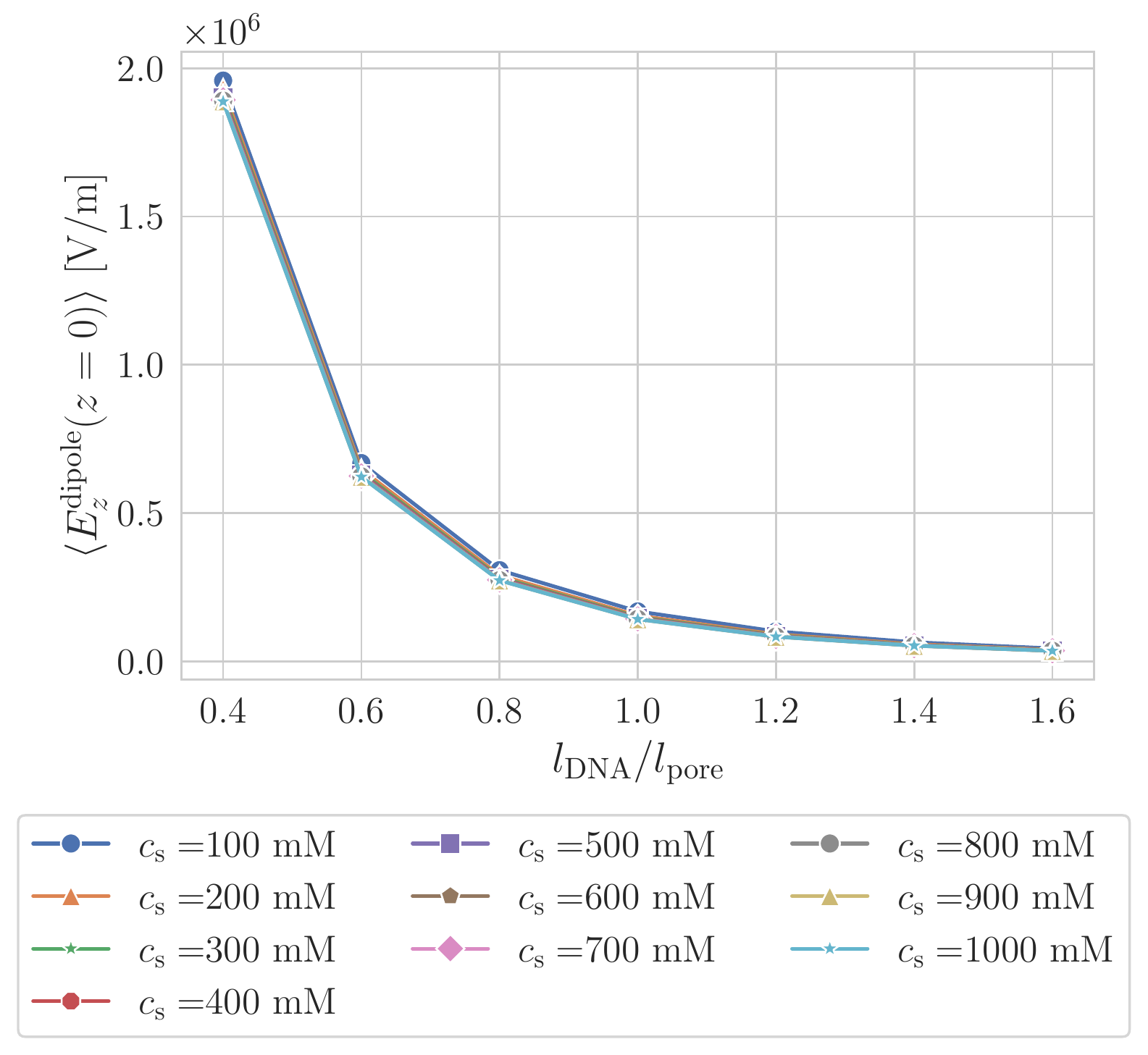}	
		\caption{The average $z$-component of the electric field caused by a single}
		\label{fig:dipole_electric_field}
\end{figure}

Thus, the electrostatic potential of these two charges ($+q$ at
$z=-0.5~l_\mathrm{DNA}$ and $-q$ at $z=0.5~l_\mathrm{DNA}$) reads:
\begin{equation}
	\begin{split}
		\Phi(r, z) = \frac{\left|q\right|}{4\pi\epsilon_0\epsilon_\mathrm{r}}&\left[\left(r^2+\left(z+\frac{l_\mathrm{DNA}}{2}\right)^2\right)^{-\frac{1}{2}} \right.-\\
		&\left. \left(r^2+\left(z-\frac{l_\mathrm{DNA}}{2}\right)^2\right)^{-\frac{1}{2}}\right].
	\end{split}
\end{equation}

From the electrostatic potential, the $z$-component of the electric field at $z=0$ can
be calculated by taking the negative gradient of the potential:
\begin{align}
	\begin{split}
		E_z^\mathrm{dipole}(r, z=0) &=\left(-\nabla \Phi(r,z)\right)_{z,z=0}\\
		&= \frac{2\left|q\right|l_\mathrm{DNA}}{\pi\epsilon_0\epsilon_\mathrm{r}\left(l_\mathrm{DNA}^2+4r^2\right)^\frac{3}{2}}.
	\end{split}
\end{align}
We now insert the assumption that the system's total dipole
moment is only caused by the two charges, \textit{i.\,e.} $\left|q\right|=\frac{\left|P_z\right|}{l_\mathrm{DNA}}$
\begin{align}
	E_z^\mathrm{dipole}(r, z=0) = \frac{2\left|P_z\right|}{\pi\epsilon_0\epsilon_\mathrm{r}\left(l_\mathrm{DNA}^2+4r^2\right)^\frac{3}{2}}.
\end{align}

\section{FEM Mesh}\label{sec:fem_mesh}

\begin{figure}[ht]
	\centering
	\includegraphics[width=0.9\linewidth]{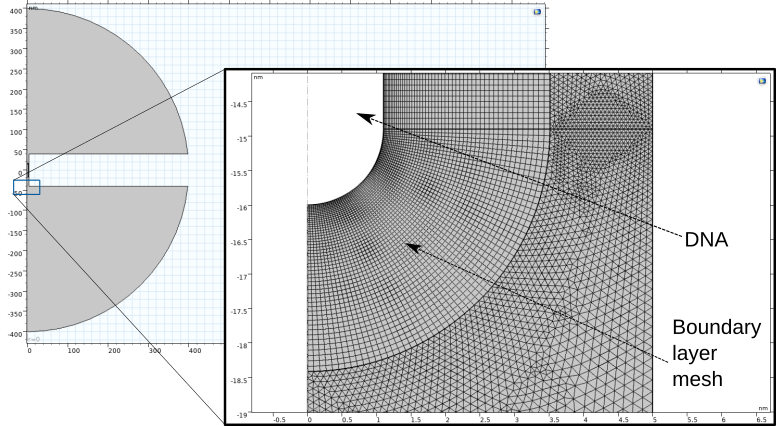}
	\caption{Example of the customized mesh used for the FEM simulations. The area around the electrically charged DNA surface is resolved with quadrilateral elements to properly capture the behavior in the Debye layer. The rest of the simulation domain consists of triangular mesh elements.}
	\label{fig:mesh}
\end{figure}

The rotational symmetry of the investigated three-dimensional system setup allows for its treatment using a quasi-two-dimensional axisymmetric simulation domain.
The FEM mesh for this domain was adapted (\textit{cf.} Fig.~\ref{fig:mesh}) in order to properly resolve the Debye layer near the negatively
charged DNA.
Using quadrilateral elements close to the DNA surface, the geometry of the mesh more closely follows the expected
symmetry of the solution for the Poisson equation for electrostatics which
defines the smallest length scale in the investigated DNA nanopore system.

\bibliography{bibtex/icp.bib}
\end{document}